\documentclass[]{jetp}

\usepackage{graphicx}

\newcommand{\lsim}   {\mathrel{\mathop{\kern 0pt \rlap
{\raise.2ex\hbox{$<$}}} \lower.9ex\hbox{\kern-.190em $\sim$}}}
\newcommand{\gsim}   {\mathrel{\mathop{\kern 0pt \rlap
{\raise.2ex\hbox{$>$}}} \lower.9ex\hbox{\kern-.190em $\sim$}}}

\begin{document}
\English

\title{Towards solving the mass-composition problem in
ultra high energy cosmic rays} 

\author{Roberto}{Aloisio}
\email{roberto.aloisio@gssi.it}
\affiliation{Gran Sasso Science Institute, L'Aquila, Italy.} 
\affiliation{INFN - Laboratori Nazionali Gran Sasso, Assergi (AQ), Italy.}

\author{Veniamin}{Berezinsky}
\email{berezinsky@lngs.infn.it}
\affiliation{Gran Sasso Science Institute, L'Aquila, Italy.}
\affiliation{INFN - Laboratori Nazionali Gran Sasso, Assergi (AQ), Italy.}

\abstract{Using the Auger mass-composition analysis of ultra high energy 
cosmic rays, based on the shape-fitting of $X_{max}$ distributions 
\cite{Auger-shape2014}, we demonstrate that mass composition and energy 
spectra measured by Auger, Telescope Array and HiRes can be brought 
into good agreement. The shape-fitting analysis of $X_{max}$ distributions 
shows that the  measured sum of proton and Helium fractions, for some 
hadronic-interaction models, can saturate the total flux. Such p+He model, 
with small admixture of other light nuclei, naturally follows from cosmology 
with recombination and reheating phases. The most radical assumption of 
the presented model is the assumed unreliability of the experimental separation 
of Helium and protons, which allows to consider He/p ratio as a free parameter. 
The results presented here show that the models with  dominant p+He 
composition explain well the energy spectrum of the dip in the latest 
(2015 - 2017) data of Auger and Telescope Array, but have some 
tension at the highest  energies with the expected  Greisen-Zatsepin-Kuzmin 
cutoff. The Auger-Prime upgrade experiment has a great potential to reject or 
confirm this model.}

\maketitle

\section{\label{sec:introduction} Introduction}

Mass composition  still remains a controversial issue in
Ultra High Energy Cosmic Rays (UHECR). The three biggest
detectors, Pierre Auger  (referred here as 'Auger'), Telescope Array
(referred as 'TA') and HiRes, have obtained contradictory results
concerning mass   composition of primary particles in the energy range
(3 - 100)~EeV   (1 EeV = $1\times 10^{18}$ eV). At (1-3) EeV all three
detectors agree with light composition, protons or protons and Helium,
but  in the range (3 - 100)~EeV the Auger detector,  the
biggest one, founds a progressively heavier mass composition 
with increasing energy, while the other two detectors report the mass 
composition similar to that at lower energy.

At present there are two basic methods to study the mass
composition  of UHECR: direct measurements and indirect tests. The
more  reliable direct method is based on the observation of the
fluorescent  light produced by Extensive Air Showers (EAS) in the
atmosphere. The  indirect test is based on the signatures 
of mass composition in the primary  energy spectrum and we start with
it as more old and less  constraining.

This approach works most efficiently for protons due to their
interaction with the Cosmic Microwave Background (CMB). It results in
two very specific spectral features: the Greisen-Zatsepin-Kuzmin
(GZK) cutoff \cite{GZK} and the pair-production dip. The former  is a
sharp cutoff at the end of the spectrum, around $E \sim 50$~ EeV,  due
to the photo-pion production and the latter is a rather faint  feature
at $E \sim 1 - 30$~EeV first calculated in \cite{Blum} and  studied in
detail in \cite{BGG,BGG-PL,Aletal}. The dip is well confirmed  in the
spectra of all three detectors but its origin as the pair-production
dip $p+\gamma_{cmb} \to p +e^{-}+ e^{+}$ is now questioned by the Auger
mass composition. Before 2011 the data published by HiRes \cite{HiResGZK}
and Auger \cite{AugerGZK}, and later confirmed by TA, showed high energy 
steepening in good agreement with the predicted  GZK cutoff. Nevertheless, 
the newest data of Auger and TA, released in 2015 - 2017 seem to be in 
contradiction with this interpretation, see Fig.~ \ref{fig:AuTA2015}.

The propagation of UHE {\it nuclei} does not leave any clear signature of
the mass composition in the energy spectra. The main channel of energy
losses, that determines the spectrum of UHE nuclei, is the
photo-disintegration process on the Extragalactic Background Light
(EBL) and on CMB. This process naturally produces secondary lighter
nuclei, mixing thus with the primary composition. As was first
predicted  by Gerasimova and Rozental \cite{GR} in 1961, i.e. before
the discovery of CMB, nuclei photo-disintegration on EBL results in a 
suppression of the UHECR energy spectrum (GR steepening). In fact, as was 
realised later, see e.g. \cite{ABG}, a more sharp cutoff occurs at
higher energies where the nucleus photo-disintegration time on 
CMB becomes equal to  that on EBL. This cutoff arises at Lorenz-factor  
$\Gamma \sim (3-5) \times 10^9$ for all nuclei. The energy of the cutoff 
$E_{cut} \propto A\Gamma$ is different for primary nuclei with different
$A$. This fact together with the unavoidable mixed composition, 
due to the production of secondary nuclei makes
unclear any composition signature in the observed spectrum.

At present the best method to measure the mass composition  is
given by the observation of fluorescent light produced by the e-m
component of EAS in the atmosphere. All three aforementioned detectors
use this method. However, for better accuracy the  fluorescent-light
method needs additional information, which in the case  of HiRes is
given by the stereo observation of fluorescent light, and  in the case
of Auger (and recently of TA) this additional information is obtained
from the data of on-ground detectors (water-Cherenkov detectors in 
Auger and scintillation detectors in TA).

The basic observable parameter related to mass composition is
$X_{\max}(E)$, the atmospheric depth where the number $N(E)$ of
particles in the cascade, with total energy $E$, reaches its
maximum $X_{\max}$, is sensitive to the number of nucleons in the
primary nucleus. Heavy nuclei interact higher in the atmosphere and
have smaller fluctuations. In practice the actual quantity which
allows to find the mass composition is the distribution $N(X_{\max})$
of the showers with total energy $E$.

In the case of large statistics the direct use of $N(X_{\max},E)$
gives the most reliable estimation of composition. In the case of
limited statistics one may use the moments of this distribution, see
e.g  \cite{Auger-moments2014}, namely the first moment which is the
mean value $\langle X_{\max} \rangle$ and the second moment
$\sigma(X_{\max})$ which is the variance or dispersion (RMS) of the
distribution. As was demonstrated in  \cite{Auger-shape2014} using
only the first two moments for the analysis, may result  in a
false degeneracy: two different mass compositions may produce the same
$\langle X_{\max} \rangle$ and $\sigma(X_{\max})$.

The shape-fitting analysis of $N(X_{max})$ recently performed by
the Auger collaboration \cite{Auger-shape2014} gives very important
results that, summarising, can be described as follows. The mass
composition is assumed as a discrete sum of four elements: Iron
(Fe), Nitrogen (N), Helium (He) and protons (p).  For each element
the $X_{\max}$ distribution is calculated by Monte Carlo (MC)
simulations and the fraction of each element in the total flux is
found from the comparison with observations. These fractions depend on
the models of hadronic interaction included in MC simulations.  A decisive
result is given by the very small fraction of Iron at all energies,
almost  independently of the hadronic interaction model (see the
upper panel  of Fig.~4 in \cite{Auger-shape2014}).
Besides, the analysis of \cite{Auger-shape2014} shows that
the fraction of light elements (p+He) is quite large independently of
the hadronic interaction model. It allows the conclusion that
at least a large fraction of UHECR, if not the dominant one, is
composed by light elements. The small fraction of Iron and large of p+He
seem to be a common conclusion of the $N(X_{max})$ shape-fitting 
analysis of Auger and HiRes/TA data.

The argument above does not dismiss the question: Auger, HiRes and TA
use the same fluorescent data to measure $X_{\max}$ and the same
moments-based method for the analysis of mass composition. Why then their
conclusions  differ?  The most convincing answer to this question is
probably given in a recent paper by the TA collaboration
\cite{TAhybrid2015}. The observation of fluorescence light can be
performed in two ways: with a monocular observation, when only one
telescope observes the fluorescent signal, or in the stereo mode when
more than one telescope simultaneously observe the same shower.
Fluorescence detection in monocular mode is less efficient to measure
$X_{\max}$ in comparison with the stereo mode. HiRes and later TA
used, apart from monocular, also stereo events with higher precision
in the measurement of $X_{\max}$. It became possible because of the
smaller (in comparison with Auger) spatial separation between
telescopes.   Auger, on the other side, to cover a larger area has a
much larger  separation among telescopes and collected mainly
monocular fluorescent  events. Instead, the Auger collaboration elaborated the 
hybrid  technique, first proposed in \cite{MIA}, based on additional accompanying
signal from, at least, one on-ground water-Cherenkov detector. Hybrid
method allows to measure the core location and geometry of the shower,
which improve the  measurement precision for $X_{\max}$ and shower
energy $E$.  Auger collected now the largest number of  hybrid events 
and we compare our predictions with the hybrid Auger data whenever this 
is possible.

At present TA is also using the hybrid technique with the help of 507
on-ground scintillation detectors \cite{TAhybrid2015}. With an accumulated 
statistics of 5 years data, TA reports \cite{TAhybrid2015} that the hybrid 
measurements of $X_{\max}$  agree with the results of Auger, if analysed with 
the EPOS-LHC  hadronic interaction model \cite{EPOS-LHC}. On the other 
hand, using the QGSJetII-03 hadronic interaction model \cite{QGSJET} the TA
collaboration founds a mass composition compatible with only light
nuclei.

Another important method to measure the mass composition is given
by the observation of muons produced in the EAS.  

The basic effect to distinguish a nucleus from a proton with the help of 
muons is related to the different energy per nucleon, $E/A$, at
fixed total energy $E$.  A low energy nucleon produces low energy
charged pions which decay to  muons before the parent pion undergoes
new collisions with air-nucleus. Produced in the EAS, muons propagate 
rectilinearly with velocity $v \approx c$. As a result they can provide directional 
and timing informations, which can further reduce uncertainties in the fluorescent
method. There are two well known muon quantities relevant for
measuring mass composition: The total number of muons in the shower 
$N_{\mu}$ and the so-called Muon Production Depth (MPD) $X_{\max}^{\mu}$,
which gives the atmospheric depth where the production rate of muons reaches 
its maximum \cite{MPR-GG2011,MPR-RC2013,mu-prod-rate}. 

The total number of muons $N_{\mu}$, called also the muon size of the
shower, is especially important to determine mass composition
at energies below the UHECR regime $<10^{17}$ eV, where the fluorescent emission 
is too faint to be detected. At these energies, $N_\mu$ gives the only way to determine
the mass composition. The analysis of the muon component
in the KASCADE-Grande experiment  \cite{KGknee} allowed recently to
find the Iron knee in the Galactic cosmic ray spectrum at energy 
$E \sim 1\times 10^{17}$~eV as predicted by the rigidity relation
\cite{Book}.

Among the three biggest UHECR arrays at present only the Auger experiment 
has several unique possibilities to measure the muon flux directly and use it to 
determine the mass composition. The on-ground  water-Cherenkov detectors can 
measure muons in inclined directions, although with a high level of uncertainty due 
to the decoupling of the electron and muon signals. In AMIGA (Auger Muon and 
Infill Ground Array) there are muon detectors in the form of scintillation counters 
buried at a depth of 2.3m underground \cite{AMIGA2015}. 

The new exciting method of muon detection in Auger experiment is given by 
the Auger-Prime \cite{AugerPrime} upgrade, recently funded, has been specifically  
designed to improve  muon detection in the whole energy range  of the experiment. 
Each water-Cherenkov tank will be equipped with scintillator layer on the top, sensitive 
only to e-m component of the shower, while  water-Cherenkov detector is sensitive to 
both e-m and muon components. The combination of the two signals allows to 
reconstruct each of the fluxes separately. Recently, also the TA collaboration started important 
upgrades to increase the statistics at the highest energies enlarging the area covered by the 
surface detectors, an updated description of the status of the TA experimental set-up can be found in \cite{TAstatus}. 

Another important channel of measuring the mass composition is connected with the 
correlation between muon signal and $X_{\max}$. We will discuss this method in some 
details in Section \ref{sec:correlation}. 

Above we discussed the problem of measuring the mass
composition. Somewhat different question is whether it is possible to
exclude, on the basis of observations, a pure proton composition at $E
> 3$~EeV. There are two challengers for such task: the quasi-isotropic
gamma-radiation and neutrinos; both produced mainly by collisions of 
UHECR protons with background photons. The most stringent limit on the
isotropic component of gamma-radiation in the range 50~MeV - 820~GeV
is given by the Fermi-LAT experiment \cite{Fermi-LAT}. The strongest
upper limit on the allowed UHE proton flux was  obtained practically
simultaneously in 2016 in three works \cite{Felix,Gavish,BGK}. The
limit depends on the models for sources of UHECR, especially strongly
on the injection power-law index $\gamma_g$ and cosmological evolution
of sources. In \cite{BGK} it was demonstrated that in a wide range of
parameters the proton models which explain UHECR flux and spectrum,
predict gamma-rays and neutrinos below the Fermi-LAT upper limit and
IceCube flux.
  
In the present paper we use the latest Auger and TA observations,
comparing them with the spectral features that arise due to
propagation  of UHECR, and their mass composition.

We argue that the spectral features may still be considered as an
indication for a light mass composition, solving the problem of the
alleged discrepancy between Auger and TA observations. 

The paper is organised as follows: In Section \ref{sec:dip} we reconsider 
the status  of the  dip model in  light of the latest observations of the
spectrum.  In Section \ref{sec:p+He} we show how a mixture of Helium
nuclei and protons  provide a good description of the observed
flux. In Section  \ref{sec:correlation} we discuss the correlation of
$X_{\max}$ with  muon characteristics: $N_{\mu}$, the total number of
muons in a shower, and $X_{\mu}^{\max}$, the Muon Production Depth;
for this discussion we calculated the spectrum in the model
p+He+CNO. The conclusion is given  in Section \ref{sec:conclusion}.

\section{\label{sec:dip} Modification factor as indication of 
proton-dominated composition}

\begin{figure}
\centering \includegraphics[scale=0.65]{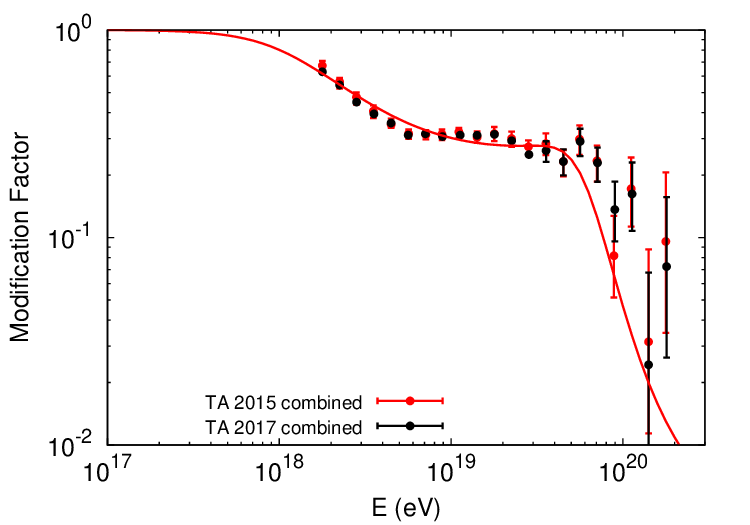}
\centering \includegraphics[scale=0.65]{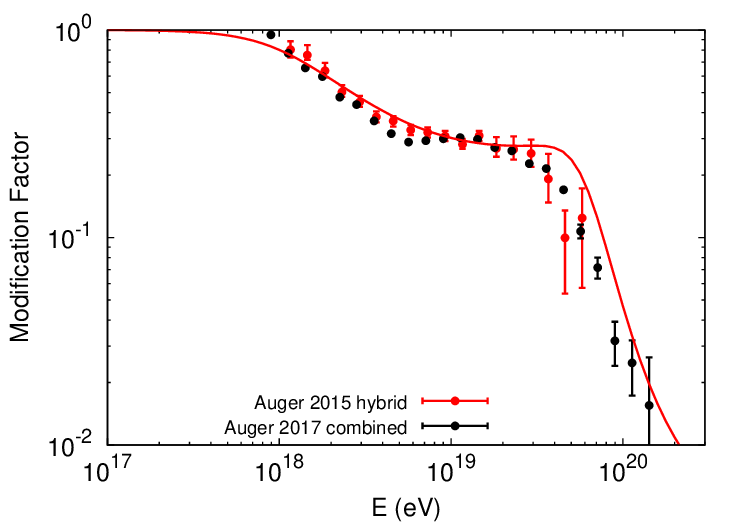}
\caption{\label{fig:AuTA2015} Proton theoretical  modification factors (red curves), 
in the case (i) $\delta E/E << \delta J/J$, compared with 2015 and 2017 combined spectra 
of TA (left panel) and of Auger 2015 hybrid spectrum and 2017 combined spectrum 
(right panel). The theoretical modification factor is computed assuming pure protons, 
with no cosmological evolution of the sources, an injection power-law index 
$\gamma_g=2.6$ with maximum acceleration energy $E_{max}=10^{21}$ eV.}
\end{figure}

Propagating through CMB the proton energy spectrum acquires two
characteristic features: the dip, due to the process $p+\gamma_{\rm
cmb} \to p+e^-+e^+$ \cite{Blum}, and GZK cutoff \cite{GZK}  due to the
reaction $p+\gamma_{\rm cmb} \to N + \pi$. These two features are
quite different from the spectral features arising in the flux 
of UHE nuclei due to the interaction with CMB and EBL. 
This difference can be used to distinguish protons 
from nuclei and can be used as an additional (indirect) test of the mass 
composition, to be compared with other observations.

This test becomes particularly important in the light of the
uncertainties in the direct measurements of the mass composition and
in the hadronic interaction models at energies above the LHC (CERN)
calibrations.

In this section we use the modification-factor method to identify
protons in UHECR. Following the works \cite{BGG,BGG-PL,Aletal}, it can
be proved that this method favours a proton-dominated mass composition
in the observations of four experiments: AGASA, Yakutsk, HiRes and
Auger, using the data before 2009, and the TA data as of 2011. 
Here we reconsider this analysis using the higher statistics data of Auger and TA 
as published in 2015-2017 \cite{Auger2015,TA2015,Auger2017,TA2017}. We will 
demonstrate that new data may agree at some reasonable conditions with the proton dip, 
but show noticeable differences with the GZK cutoff (see Fig.~\ref{fig:AuTA2015}).

The proton modification factor used in the forthcoming analysis is defined as the ratio:  
\begin{equation}
\eta_p(E)=\frac{J_p(E)}{J_p^{\rm unm}(E)},
\label{eq:eta}
\end{equation}
where $J_p(E)$ is the total flux of protons, measured or
calculated, taking into account all energy losses due to 
$p\gamma_{\rm cmb}$ collisions and adiabatic energy losses. 
In Eq.(\ref{eq:eta}) we introduced also the unmodified proton spectrum 
$J_p^{\rm unm}(E)$ which is calculated taking into account only adiabatic 
energy losses, $J_p^{\rm unm}(E)=KE^{-\gamma_g}$.

{\em Model-dependent phenomena} enter both numerator and denominator in
Eq.(\ref{eq:eta}) and compensate or even cancel each other, while
interaction with CMB photons does not enter the unmodified flux,
i.e. the denominator, and appears only in the numerator of
Eq.(\ref{eq:eta}). Thus the modification factor presents, in an
unsuppressed way, such features as dip and GZK cutoff directly connected
with the propagation of UHE protons, while the model-dependent features are 
seen there in a suppressed form. Therefore the modification factor is
an excellent instrument to search for the proton-dominated mass
composition through the proton interaction features, dip and GZK cutoff,
but it is not sensitive to the model-dependent features e.g. to the details of the 
acceleration models.

Modification factor depends non-trivially on the statistics of the events.

First, at very high observational statistics the agreement of the observed modification 
factor with the predicted one (both for protons) must be worse because at high resolution 
the ratio starts to distinguish better the model-dependent effects being less compensated 
in denominator and nominator of Eq.(\ref{eq:eta}). In other words the higher statistics 
results in a worse agreement between denominator and numerator in  Eq.(\ref{eq:eta}) 
due to model-dependent effects. 
This phenomenon  will be referred to as "{\bf high-statistics de-compensation}". 

The second effect is caused by the observation technique. There are two kinds of observational 
errors: flux errors $\delta J/J$ and energy errors $\delta E/E$. In all published
presentations of the measured spectra the errors are given as the flux errors $\delta J/J$, 
while the energy errors $\delta E/E$ are just mentioned or shortly discussed. 
The systematic energy error was first presented in the Auger ICRC report 2013 
\cite{Verzi2013,Rapporteur2015} as $\delta E/E = 0.14$.
 
In the present paper we consider two cases:\\
 
\noindent (i) The total error is  dominated by flux error 
$\delta J /J >> \delta E/E$, \\ 

\noindent (ii) The energy error $\delta E /E$ is the dominant component 
 $\delta E/E > \delta J/J$, or  $\delta E/E >> \delta J/J$.\\*[2mm]
 
We start with the modification factor calculation $\eta (E_{\rm obs}) \pm \delta \eta$ in the 
general case of two errors $\delta E_{\rm obs}$ and $\delta J_{\rm obs}$. It is easy to obtain 
the exact expression for $\eta(E_{\rm obs})$  in the case of natural condition 
$\delta E_{\rm obs} <<  E_{\rm obs}$:
\begin{equation} 
\eta(E_{\rm obs}) \pm \delta \eta = 
\frac{J_{\rm obs}(E_{\rm obs})} {K E_{\rm obs}^{-\gamma_g }} \times
\left[1 \pm \frac{1}{J_{\rm obs}(E_{\rm obs})} 
\frac{\partial J_{\rm obs}(E_{\rm obs})}{\partial E_{\rm obs}}\delta E_{\rm obs} 
\pm \frac {\delta J_{\rm obs}(E_{\rm obs})} {J_{\rm obs}(E_{\rm obs})} \right]
\label{eq:theor-errors}
\end{equation}

\begin{figure}
\centering \includegraphics[scale=0.65]{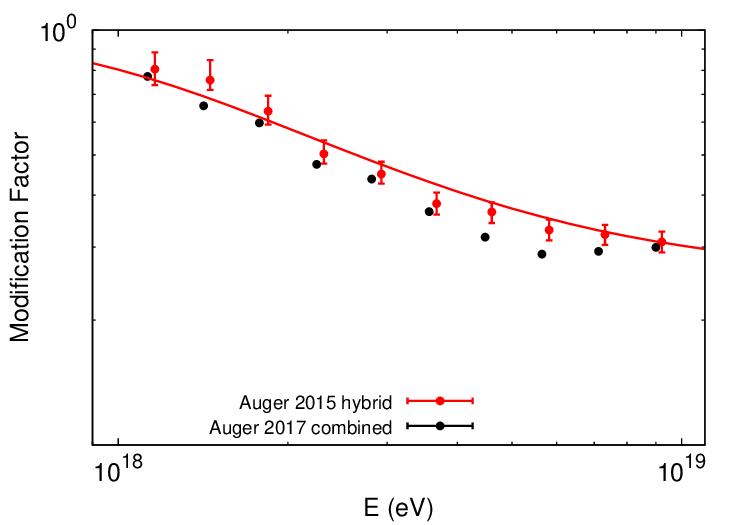}
\centering \includegraphics[scale=0.65]{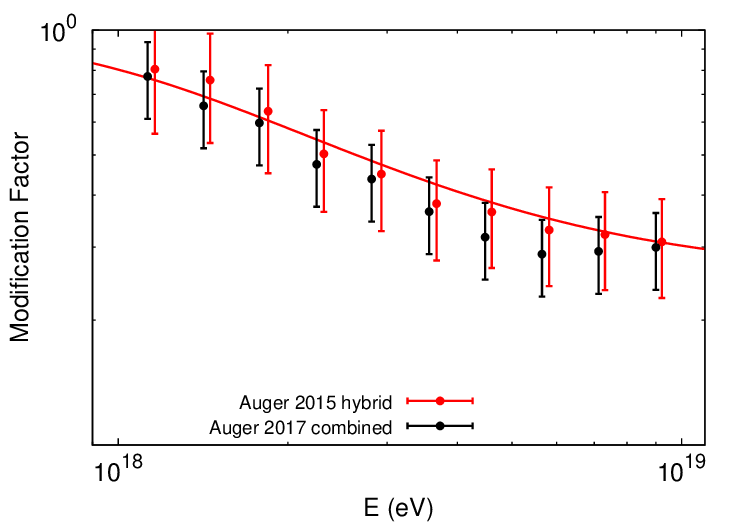}
\caption{\label{fig:allerr}
{\bf [Left panel]} Auger modification factor in the energy range 
1 - 10 EeV obtained from the hybrid spectrum 2015 (red points) and 
from the combined spectrum 2017 (black points) in comparison with 
the theoretical modification factor (solid curve). In the energy interval (4 - 6) 
EeV one may notice the difference between hybrid and combined data and the 
difference of both with the theoretical modification factor (the gap). 
{\bf [Right panel]} The same data as in the left panel, with energy 
error $\delta E$ taken as a half of the systematic energy error of Auger 
\cite{Verzi2013,Rapporteur2015}. One may notice that the gap between
hybrid and combined data, and between data and theoretical 
modification factor are over-closed by the energy error-bars.}
\end{figure}

In the case (i), $(\delta E/E)_{\rm obs} <<(\delta J/J)_{\rm obs}$, the 
second term in rhs of Eq.(\ref{eq:theor-errors}) may be neglected and 
the comparison of the proton modification factor with the data of four
experiments released before 2011, namely AGASA, Yakutsk, HiRes, Auger,
and also the 2011 data of TA, results in an excellent agreement with the 
observed spectra in approximately 100 energy bins \cite{BGG,BGG-PL,Aletal}. 
It is a remarkable fact that this agreement is achieved with only one
free parameter, the injection power-law index $\gamma_g \approx 2.6$,
together with sources emissivity ${\cal L}_0$, which provides the
total normalization of the flux using the same restriction
$(i)$ ~  $\delta E/E < \delta J/J$.

In Fig. \ref{fig:AuTA2015} we compare the theoretical modification factor for 
protons with TA (left panel) Auger (right panel) data as of 2015 and 2017, using the same restriction 
$ (i) ~  (\delta E/E) < \delta J/J$. This comparison shows a fairly good agreement with the dip in both datasets, 
the relatively small discrepancy with GZK cutoff for Auger and stronger discrepancy with GZK cutoff for TA. 
We postpone the discussion of GZK cutoff to a forthcoming publication. 

As to more details concerning the dip, TA combined spectra of 2015 and 2017, show an excellent agreement 
with the theoretical modification factor, while Auger data show a good agreement with theoretical curve for 2015 
hybrid data, but for the high-statistics 2017 combined data there is statistically significant difference with 
both the theoretical spectrum and the 2015 hybrid spectrum. Since the statistics of the combined Auger 
events is much higher than that for TA and for hybrid Auger events 2015, one may suspect 
the "statistical de-compensation" effect as the explanation for discrepancy, namely suppression of the 
compensation mechanism for model-dependent effects in numerator and denominator of modification 
factor given by Eq.~ (\ref{eq:eta}).

The reasonable agreement of the modification factor calculated above within the assumption 
$\delta E /E << \delta J/J$ may be considered as experimental confirmation of this assumption, 
however there is an argument in favour of alternative  relation (ii) $\delta E/E > \delta J/J$. 
From the Auger presentations of the combined energy spectra of 2014 - 2017 in the form 
$E^3J(E)$ with error-bar $\delta\left[E^3J(E)\right]$ :
\begin{equation} 
\frac{\delta(E^3J(E))}{E^3 J(E)} = \frac{\delta J(E)}{J(E)}\\ 
+ 3 \frac{\delta E}{E} ,
\label{eq:sum-errors}
\end{equation}
it is important to note that the total error $\delta(E^3J(E))$ is determined by the errors $\delta J$ 
and $\delta E$, which operate in two perpendicular directions. This remark is also valid for the 
modification factor $\eta(E)\propto E^{\gamma_g} J(E)$, which results in 
\begin{equation}
\frac{\delta\eta(E)}{\eta(E)}=\frac{\delta J(E)}{J(E)} + \gamma_g\frac{\delta E}{E}~.
\end{equation}

One may conclude  that $\delta E/E$ term can be larger or much larger than $\delta J/J$  
in the case of high statistics Auger combined events. Indeed, the first term $\delta J/J$ 
may be very small because of the tremendous Auger statistics (especially for the combined events) 
while the second term is large in particular for the systematic errors 
$\delta E/E=0.14$ \cite{Verzi2013}. Moreover, the difference between the hybrid and 
combined spectra in the Auger data (see below) could point toward a higher energy 
error. However, it is difficult to estimate it and we restrict our analysis to 
what discussed in \cite{Verzi2013}. Thus $\delta E/E \gg \delta J/J$, 
and it may over-close the whole dip in the vertical direction making it unobservable: 
the dip exists but cannot be seen because of too large energy errors. We will refer 
to this effect as {\bf "over-closing of gap"}.

This case is illustrated by Fig.~\ref{fig:allerr} where the theoretical modification factor is plotted 
together with the Auger combined spectrum of 2017 and Auger hybrid spectrum of 2015. 
The two spectra show the largest difference in the recent Auger data. 
In the left panel they are presented with error bars $\delta J/J$, and 
in the right panel the energy errors $\delta E/E$ are added as half of 
systematic energy error of Auger \cite{Verzi2013,Rapporteur2015}.   
One may see that energy error-bars over-close the gap between the two Auger 
spectra and between each of Auger spectrum and theoretical modification 
factor (see  captions to Fig.~\ref{fig:allerr}).  

\begin{figure}
\centering
\includegraphics[width=\textwidth]{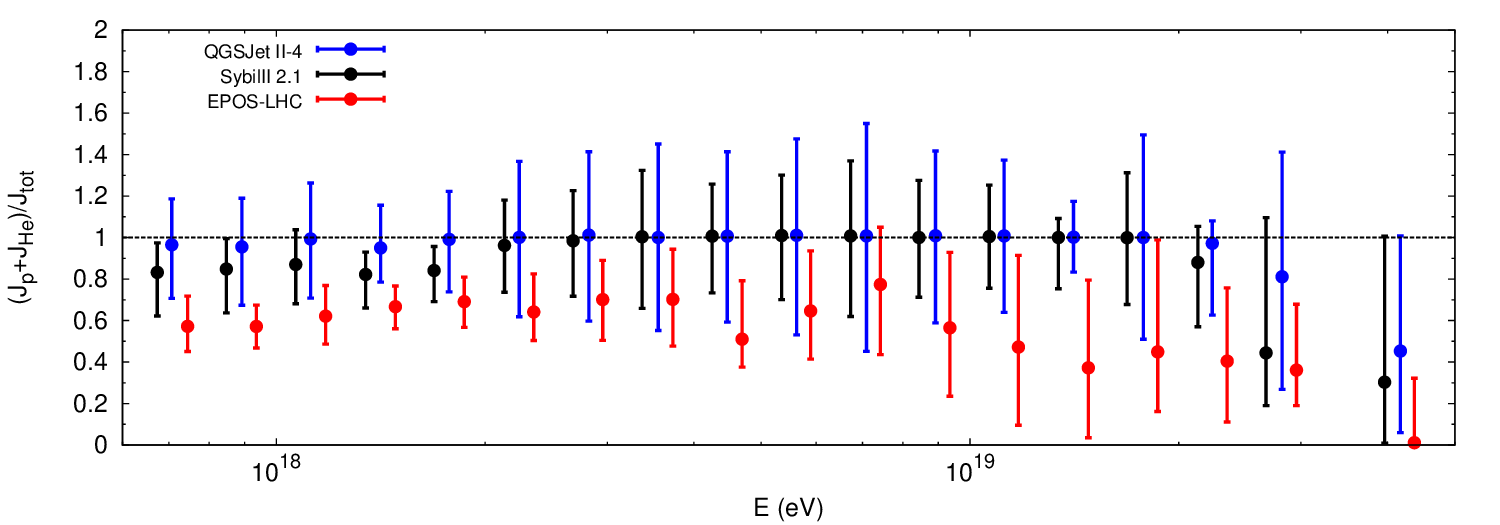}
\caption{Fraction of p+He nuclei relative to the total one according to the 
Auger measurements \cite{Auger-shape2014} with error bars
(summed in quadrature) taken from Fig.~4 of \cite{Auger-shape2014} using hadronic 
interaction models: Sibyll2.1 (black), QGSJETII-4 (blue) and EPOS-LHC (red).}
\label{fig:p+He}
\end{figure}

The conclusion above is valid for small energy errors
$\delta E/E < \delta J/J$. In the case of larger energy errors 
$\delta E/E > \delta J/J$ the energy error $\delta E$ can exceed the gap between predicted 
and observed values, $\eta_{th}$ and $\eta_{obs}$, in particular at most disturbing 
energies $E\sim 4 - 6$~EeV, and contradiction between $\eta_{th} (E)$ and $\eta_{obs} (E)$ 
disappears.

In a more general discussion the good agreement of the proton theoretical
modification factor with Auger and TA observations, at least at the
dip energies $1 - 30$ EeV, is a strong indication of proton or
proton-dominated composition. However, one cannot consider it as the
the final proof. Indeed, on one side we observe the unique shape of
the dip produced by $p+\gamma_{\rm cmb} \to p+e^-+e^+$ scattering. On
the other side, in models with mixed nuclei composition, see for
instance \cite{ABB}, one can obtain a theoretical spectrum with
practically the same shape of the pair-production dip, but using more
than 10 free parameters in the theoretical model. This result
demonstrates that the very specific shape of the pair-production dip
is not the unique one explaining the observed spectrum. 

As discussed in the Introduction, there are a few other observations
that can in principle challenge the proton-dominated mass composition:
the Auger observations of mass composition using fluorescence and
muon signals, the Fermi LAT data for the diffuse gamma-ray background
and the IceCube data for astrophysical neutrinos. If all these
experiments provide in future evidences against the
proton-dominated composition one must conclude that the
pair-production dip is an accidental coincidence. Nevertheless, until
experimental data are not conclusive, one must consider the
proton modification factor as an indication for a
proton-dominated mass composition of UHECR.

\section{\label{sec:p+He} $p+He$ model}

As discussed in the Introduction there is some tension on   
the mass composition at $E> 3$~ EeV between the three biggest 
UHECR experiments Auger, TA and HiRes. In this Section we first summarise 
the basic physics of mass-composition measurements, then discuss the 
influence of cosmological environment and, finally, present calculations relevant 
to the $p+He$ model.

The measurement of mass composition is based on the $X_{\max}$ value,
which is the depth of the atmosphere where the number of particles in
the shower reaches its maximum. The value of $X_{\max}$ is a basic
parameter to determine the mass composition of UHECR, with the best
observable quantity for this determination given by the shape  of
the distribution $N(X_{\max},E)$ for  showers with fixed total energy
$E$.

As a matter of fact, until recently, instead of the distribution
$N(X_{\max})$, the first two moments of this distribution were used:
the mean value $\langle X_{\max} \rangle$ and its RMS 
$\sigma(X_{\max})$.

In two recent papers by the Auger collaboration, the mass-compositions 
obtained using moments-analysis \cite{Auger-moments2014} and shape-fitting
$N(X_{max})$ analysis \cite{Auger-shape2014} are not identical. Realistically, 
they are not expected to be such,  similarly to the already known fact 
that $\langle X_{\max} \rangle$ and $\sigma (X_{\max})$ give, if analysed separately, 
somewhat different results. The shape-fitting analysis is obviously the most
fundamental and most sensitive method, since, for example, it involves
the tiny parts of the wing distribution. Apart from it, the
shape-fitting analysis demonstrated a {\em degeneracy effect} when
two different mass compositions correspond to the same first two
moments. For this reason, in the present paper, we choose the results
obtained from the shape-fitting analysis of the Auger data
\cite{Auger-shape2014} with the measured fractions of four nuclei
species: Fe, N, He and p. These fractions, as determined from Auger
measurements \cite{Auger-shape2014}, reveal some uncertainties due to
different hadronic interaction models namely QGSJet \cite{QGSJET},
EPOS-LHC \cite{EPOS-LHC} and Sybill \cite{Sibyll}.

The important result obtained in \cite{Auger-shape2014} is given by a
very small fraction of Iron at all energies and for all interaction
models, except EPOS-LHC at the two highest energy bins (see the upper
panel in Fig.~4 of \cite{Auger-shape2014}). The other important
result, as mentioned in Introduction, and exposed in 
Fig.~\ref{fig:p+He}, is a large fraction of p+He, consistent with
unity.  

\begin{figure}
\centering \includegraphics[scale=0.65]{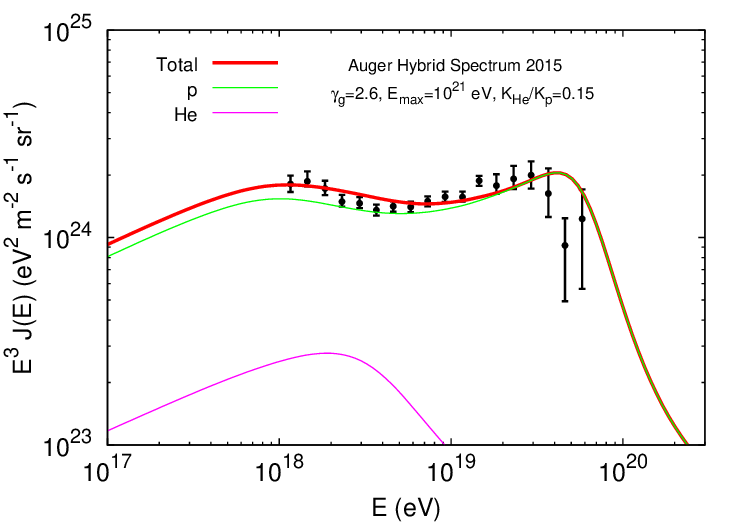}
\centering \includegraphics[scale=0.65]{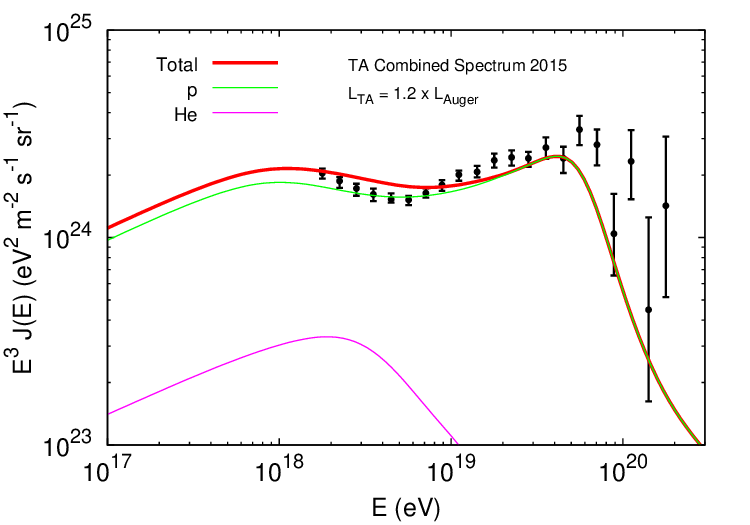}
\caption{{\bf [Left panel]} Energy spectrum for mixed p+He composition 
with injection index $\gamma=2.6$, $E_{max}^{acc}=10^{21}$ eV and $He/p=0.15$ 
in the generation spectrum. The contribution of He is shown by magenta curves, 
proton by green and total by red. The calculated spectrum is compared with hybrid data 
of Auger 2015. {\bf [Right panel]} As in left panel in comparison with TA combined
data. The agreement needs a rescaling of the total flux by a factor
1.2. One may notice the excess of TA flux at energy above the GZK cutoff.}
\label{fig:pHeSpectr_g26}
\end{figure}

It is interesting to note that both effects have a natural
cosmological explanation.    
  
Among the heaviest nuclei, Iron is the most natural element to be
produced in Supernova (SN) explosions and the absence of Iron in
UHECR implies that other heavy elements must be absent too.  Their
suppression in the form $ Fe/p \ll 1$  is very natural for 
extragalactic gas and extragalactic cosmic rays. Enhancement of 
p+He component has the same nature.

At the cosmological epoch of {\it recombination}, protons and Helium 
nuclei were the dominant components and heavy metals were almost 
completely absent. Meanwhile, the  production of metals is compulsory. 
It is needed to provide cooling of ordinary stars during their evolution, 
including the preSN phase. The later stage of {\it reionization}
in the universe, as detected by WMAP \cite{WMAP} and  Planck \cite{Planck} 
satellites occurs at redshift $z=11.0 \pm 1.4$ and $z\lesssim 10.0$, respectively. 
This stage needs at least two early generations of stars with low metallicity, 
Pop III and Pop II stars. They inject into the extragalactic space a small amount of 
heavy metals. The main contribution to the Iron observed in the extragalactic 
space (and thus in extragalactic cosmic rays) is given by the present-time SN 
explosions. This scenario  is confirmed by WMAP and Planck observations of the 
Universe reionization and by the observations of  $Ly \alpha$ forest which 
indicate that extragalactic space had very low fraction of heavy elements at 
the level $Z \sim 10^{-3.5} Z_\odot$ at redshift $z \sim 5$, see eg. \cite{lowFe}. 
Iron and other heavy metals are injected into  extragalactic space mainly during 
a short interval $\Delta z$ at $z \sim 0$ mostly due to explosions of the last 
generation of SNe. This scenario is similar to the model of UHECR 
produced mostly nearby our Galaxy \cite{aharonian}.

One may conclude that Hydrogen and Helium as the main products of  
primordial cosmological nucleosynthesis, and suppression of SN-produced 
Iron and other heavy metals in red-shifted gas, naturally result in 
a p+He dominated extragalactic gas and UHECR accelerated at red-shift 
$z\gsim afew$.

In our calculations an additional simplifying assumption is
used. Generically, we assume that all existing detectors do not
distinguish reliably Helium from proton and one can consider p+He flux
as one light component, assuming the fraction $He/p$ as a free
parameter  of the model. However, we will start with the sum p+He as
it comes from Fig. 4 (strip 3 for He and strip 4 for p) of
\cite{Auger-shape2014}. Summing these two fractions, with errors
summed in quadrature, we obtain  p+He flux presented in
Fig.~\ref{fig:p+He}. One can see that the sum of these two components
saturates well the total flux, at least in the case of QGSJet and
Sybil hadronic interaction models. This interesting fact confirms well
our assumption that the light fraction (p+He) weakly depends on energy
and with good accuracy saturates the total flux leaving small room for
other components (e.g. $N$ which will be considered later as CNO
component).

\begin{figure}
\centering \includegraphics[scale=0.65]{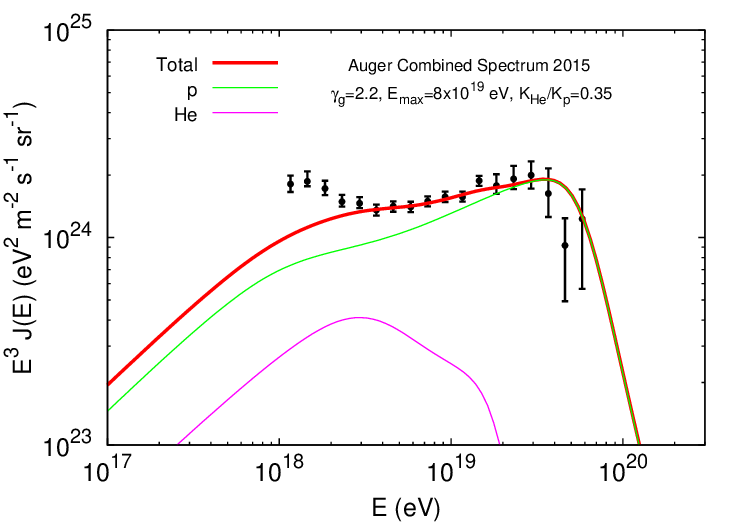}
\centering \includegraphics[scale=0.65]{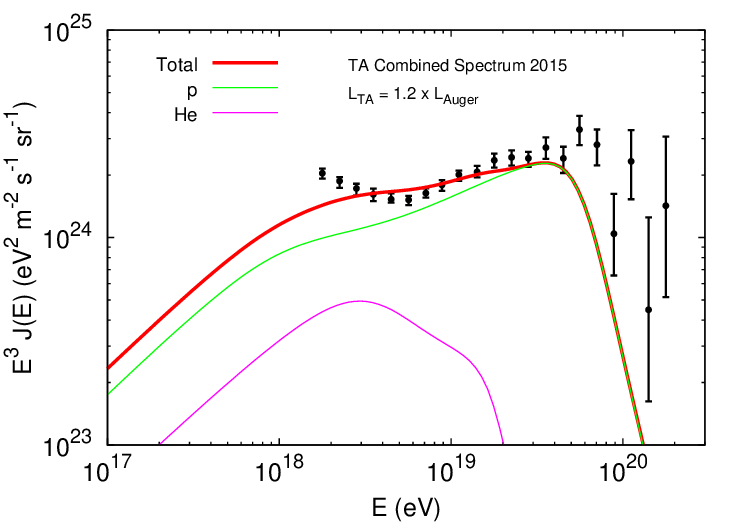}
\caption{ The same as in Fig.~\ref{fig:pHeSpectr_g26} but
for $\gamma_g=2.2$, $E_{max}=8\times 10^{19}$ eV and a ratio of
protons and Helium nuclei at the source He/p=0.35.}
\label{fig:pHeSpectr_g22}
\end{figure}

We are ready now to calculate the energy spectra for $p+He$ models and
to compare them with  spectra released by Auger and TA in 2015.  We
consider a power-law generation spectrum as $Q(E)=K_i E^{-\gamma_g}$
($i=p,~He$) with the same generation index $\gamma_g$ for protons
and Helium nuclei. We also assume that sources are distributed
homogeneously and uniformly, so that the calculated spectrum is
universal, i.e. not being affected by propagation models. Energy
losses  include pair-production, photo-pion production, and
photo-dissociation  for Helium on CMB and EBL. 

Secondary protons from $He$ and $D$
photo-dissociation and also from neutron decays are included in
calculations. For interaction with EBL photons the model \cite{EBL} is
used. In all these calculations we follow \cite{ABB}.

In Fig.~\ref{fig:pHeSpectr_g26} and Fig.~\ref{fig:pHeSpectr_g22} the
computed spectra for $p+He$ models are presented for $\gamma_g=2.6$
and  $\gamma_g=2.2$, respectively. A generic feature of $p+He$ spectra
is the proton dominance at the highest energies because of the GR
steepening for Helium at  $E\lesssim 5\times 10^{18}$ eV due to
photo-disintegration on EBL. Therefore the GZK cutoff  in $p+He$ model
becomes compulsory, unless the maximum acceleration energy
$E_{\max}^{\rm acc}$ is below the GZK threshold  $E_{\max}^{\rm GZK}
\simeq 50$ EeV.

Consider first the case of generation index $\gamma_g=2.6$  shown in
Fig.~\ref{fig:pHeSpectr_g26}. This generation index corresponds to
the canonical proton modification factor in the dip model (see section
\ref{sec:dip}). Therefore if  to take a small $He/p$ ratio at
generation one should obtain the theoretical spectrum and theoretical
modification factor in agreement  with old (before 2015) observations.
One may notice from Fig.~\ref{fig:pHeSpectr_g26} the similar
agreement between theoretical and  observational dips for Auger 2015
(hybrid data) and for TA 2015 (combined data). This agreement becomes
worse  at the highest energies.

In Fig.~\ref{fig:pHeSpectr_g26} we plot the comparison of the observed 
and calculated fluxes for Auger 2015 (left panel) and TA 2015 (right panel). 
It is remarkable that, at the dip energies, the TA spectrum can be described 
just rescaling by a factor $1.2$ the source emissivity needed to 
describe the Auger data. The behaviour of the flux at the highest energies 
is determined by the photo-pion production process. The maximum acceleration 
energy in Fig.~\ref{fig:pHeSpectr_g26} is taken at the level of $E_{max}=10^{21}$ eV. 
In other words the theoretical spectrum shape in Fig.~\ref{fig:pHeSpectr_g26} is 
exactly as predicted in the case of the GZK cutoff and, as follows from 
Fig.~\ref{fig:pHeSpectr_g26}, it seems not well reproduced in both data 
sets. Auger shows an earlier cutoff at energies below the GZK cutoff energy 
($\simeq 50$ EeV) while TA shows a flux suppression at energies slightly 
higher than this value.

The fraction of Helium allowed at the sources depends on the
assumptions for the injection power-law index. Assuming harder spectra
it is possible to increase the fraction of Helium. In
Fig.~\ref{fig:pHeSpectr_g22} we assume a maximum acceleration energy
$E_{max}=8\times 10^{19}$ eV and a flatter injection spectrum with
$\gamma_g=2.2$, that allows to increase the fraction of Helium nuclei
in  the generation spectrum up to $K_{He}/K_p=0.35$. This procedure
improved but a little the agreement  with observational data of Auger
at the highest  energies, while the good agreement with the dip
remains practically as before. These changes are linked with the
GR steepening of He spectrum  due to photo-disintegration on the EBL
radiation.
 
\section{\label{sec:correlation}$X_{\max}$ and muons}

As discussed in the Introduction, the observation of the showers muon
component is a very efficient tool to measure the mass
composition. The total number of muons  $N_{\mu}$ in the shower 
and also the Muon Production Depth $X^{\mu}_{\max}$ are sensitive  
indicators for it. These two quantities are expected to have correlations 
with $X_{\max}$, measured using the fluorescent light, because all three 
quantities characterise the mass composition. Among the three biggest 
arrays, Auger, at present, is the most efficient one to measure the muon 
flux and the correlations of $N_{\mu}$ and $X^{\mu}_{\max}$ with $X_{\max}$.

\begin{figure}
\centering \includegraphics[scale=0.65]{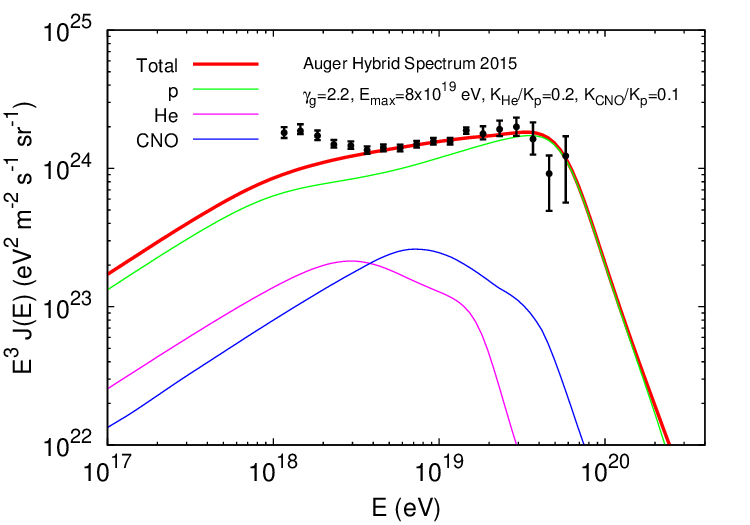}
\centering \includegraphics[scale=0.65]{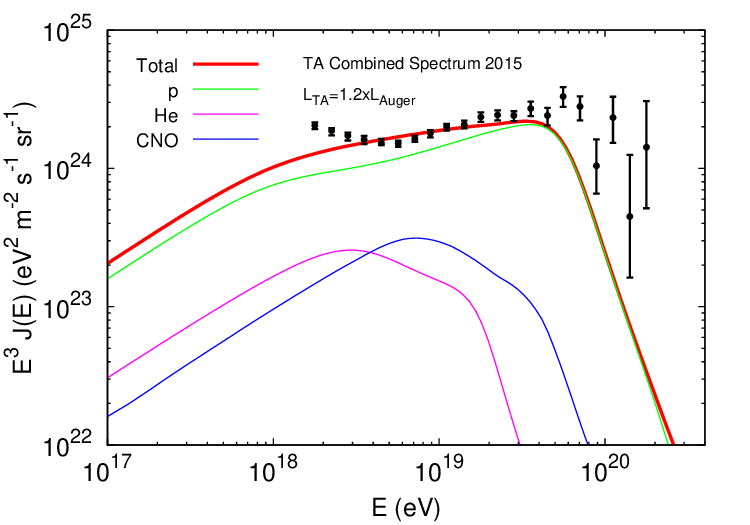}
\caption{Energy spectrum of protons, Helium and CNO with generation
ratio 1:0.2:0.1, spectral injection index $\gamma_g=2.2$ and maximum energy 
of acceleration $E_{\max}=8 \times 10^{19}$~eV. The calculations are 
compared with the 2015 release of hybrid data of Auger (left panel) and 
combined data of TA (right panel).}
\label{fig:CNO}
\end{figure}

To pursue such measurements the Auger collaboration has just started an 
overall upgrade of the experimental set-up (Auger-Prime) to instrument each 
water-Cherenkov detector with a $4m^2$ plastic scintillator located on the top 
\cite{AugerPrime}, to disentangle  the total signals produced by muons and 
electrons.

From the theoretical point of view, P. Younk and M. Risse
\cite{correl-th} were the first to demonstrate that the statistical
correlation of the shower-maximum depth $X_{\max}$ and the total number of
muons $N_\mu$ in the shower, depends on the mixture of different
nuclei species in the UHECR flux and thus can be used for the analysis
of the mass composition. This is a transparent theoretical idea,
because both quantities are good characteristics of the mass
composition measured, however, in two independent experimental
ways. In \cite{correl-th}, the authors introduced the statistical
correlation factor $r$, see Eq.(1) in \cite{correl-th}, for the two
quantities $X_{\max}$ and $N_\mu$ and demonstrated that $r$ is very
sensitive to single nuclei composition and their mixtures. For
example, a pure proton and a pure Iron composition gives $r_p=0.0$ and
$r_{\rm Fe}=0.7$ respectively, while for equal ratios of both nuclei
$r_{p+Fe}=-0.51$. These numbers are given for the case of an 'ideal
detector', for a realistic detector the difference is smaller. This
example demonstrates the power of this method to shed additional light on
the actual mass composition of UHECR.

Recently, restricting the analysis to hybrid events, the Auger
collaboration presented the measurement of the correlation between the
depth of shower maximum $X_{max}$, as observed by fluorescence
telescopes, and the signal in the water-Cherenkov detectors
\cite{correl-exp}. The surface array of these detectors is sensitive
to muons, particularly in the case of inclined showers with zenith
angles between $20$ and $90$ degrees, in which muons provide from 
$40\%$ to $90\%$ of the signal $S(1000)$ at a distance 1000 m from the 
shower core. Instead of $X_{\max}$ and  $S(1000)$  in \cite{correl-exp}   
as correlation quantities there were used $X_{\max}^*$ and $S_{38}^*$, 
which are the values of $X_{\max}$ and $S(1000)$ recalculated to a shower 
zenith angle of 38$^{\circ}$ and to a shower energy of 10~EeV. The statistical 
correlation factor measured and simulated in this way is denoted in \cite{correl-exp} 
as $r_G(X_{\max}^*,S_{38}^*)$. The measured value obtained is 
$r_G^{\rm data}=-0.125 \pm 0.024$ (negative), while the simulated values of the 
correlation factors for pure protons (p) and for a mixture of protons and Helium 
(0.8p+0.2He) were found approximately $r_G \approx 0$, for all three hadronic 
interaction models EPOS-LHC, QGSJetII-04 and Sibyll 2.1. This result seems 
to exclude both a pure proton model and a proton model with an admixture of 
20\% He, moreover, any p+He model, without an admixture of heavier nuclei, is
disfavoured. The last conclusion is a strong argument against the pair-production 
dip, though this model may survive including He and more heavy nuclei as small admixtures. 

On the other hand one should realise that the present correlation measurement 
is undoubtedly a step forward to enter the muon physics in UHECR, since the correlation 
involves muon signals. Auger \cite{muon-excess} and TA \cite{TAmuon-excess} 
face the problem here  in the form of a muon excess by a factor 1.5 - 2.0 higher then expected 
at primary energies above 10~EeV. The analysis \cite{Ostap2016} shows that this excess 
is difficult to explain by any reasonable modification of the particle interactions and it maybe 
evidence for a light (proton) composition at these energies. The authors of \cite{TAmuon-excess} 
also interpret their measurements by a proton primary composition. Moreover, the muon excess 
questions the accuracy of the measurement of the muon signals based on inclined showers, 
and requires some caution towards the mass-composition restrictions obtained in \cite{correl-exp}. 
One may hope that in the near future the Auger-Prime measurements of $N_{\mu}$ and $X^{\mu}_{\max}$ 
will allow to find the correlation factors $r(X_{\max},N_{\mu})$ and  $r(X_{\max},X^{\mu}_{\max})$  
with the needed  accuracy.  

As discussed in the Introduction, a powerful method to determine
mass composition with the help of muons is given by the measurement
of the muon production depth of the shower $X_{\max}^{\mu}$
\cite{MPR-GG2011,MPR-RC2013,mu-prod-rate}. Muons trace their parents,
heavy nuclei or protons, being produced at different heights in the
atmosphere and propagating rectilinearly with velocity $v \approx c$.
The timing of muon signals according to detailed calculations \cite{mu-prod-rate} 
will improve the accuracy of $X_{\max}^{\mu}$ method as well as the accuracy of 
its correlation with $X_{max}$.

The problem with mass composition as it is found in \cite{correl-exp} 
requires the presence of nuclei heavier than Helium. Here, assuming 
that the result of \cite{correl-exp} will be confirmed in a more convincing 
way through the correlations $(X_{\max}, X_{\max}^{\mu})$,
and $(X_{\max}, N_{\mu})$, we included in $p+He$ model also CNO nuclei
with ratios at the source $He/p$=0.2 and $CNO/p$=0.1 (both allowed 
by Auger data \cite{Auger-shape2014}). The spectra obtained are shown 
in figure \ref{fig:CNO}, calculated following the computation scheme of 
\cite{ABB}, show a quite good agreement with experimental data of both 
Auger and TA. The latter, as before, are reproduced assuming a source 
emissivity multiplied by a factor 1.2 respect to the Auger case.

\section{\label{sec:conclusion} Conclusions}

As far as mass composition is concerned there are three methods to
analyse the fluorescence data: $\langle X_{\max} \rangle$,~ $\sigma
(X_{\max})$ and the shape-fitting analysis of  $N(X_{max})$
distribution \cite{Auger-shape2014}. As it is well known, the first
two methods (moments of the $X_{\max}$ distribution) do not agree well
between themselves and both disagree with the shape-fitting analysis
(compare the mass composition obtained in \cite{Auger-moments2014} and
in \cite{Auger-shape2014}). In this paper we used the Auger
shape-fitting analysis as the most reliable and free from false
degeneracies, see \cite{Auger-shape2014} and Introduction.

In the shape-fitting analysis the mass composition is described in
terms of four nuclei species: p, He, N (we consider it as CNO) and
Fe (which can be considered as Iron group including the heavy metals). 
The results of this analysis are given as fractions of the fluxes
of these four elements, which depend rather strongly on the hadronic
interaction models used.  The new and important result of this
analysis is the very small fraction of Iron, compatible with zero,
practically for all models of hadronic interactions. We argue that
this result is natural for the standard cosmology with reionization of
the universe.

Our first observation is that using QGSJet II-4 and Sybill 2.1 for the
hadronic interaction model the sum of protons and Helium nuclei
fractions saturates with good precision the total flux, while for
EPOS-LHC  it leaves more space for other elements especially at the
lowest and highest energies. Thus a reasonable model could be a p+He
dominated injection with a  small admixture of CNO muclei.

Next we made the ad hoc assumption that at present all existing
detectors cannot distinguish reliably Helium nuclei from protons and
we calculated the spectra for Helium and protons considering them as  a
single component with the same injection power-law index $\gamma_g$
equal to 2.6 and 2.2 and taking the ratio He/p at the source to fit
the spectra of Auger and TA. These ratios are 0.15 for $\gamma_g=2.6$
and 0.35 for $\gamma_g=2.2$. The calculated spectra are shown in 
Figs.~\ref{fig:pHeSpectr_g26} and \ref{fig:pHeSpectr_g22}.

The highest energy part of these spectra are always dominated by
protons, because high-energy He nuclei are photo-disintegrated in
collisions with EBL photons. 

In the case of $\gamma_g=2.6$ the observed dip is mainly produced 
by protons: it is the canonical case of the dip model  considered in section 
\ref{sec:dip} in terms of the modification factor. The new element  of this study 
is the inclusion of two errors: flux error $\delta J/J$ and energy error $\delta E/E$. 
At very large statistics of events the energy error dominates, and using 
$\delta J/J$ error e.g. for spectrum presented in the form  $E^3 J(E)$ is 
incorrect. It changes the status of the dip model in terms of the combined 
Auger spectrum (see Section \ref{sec:dip}).

In the case $\gamma_g=2.2$ the dip at EeV energies of the spectrum is
produced by both Helium and proton components. However, the observed  
high energy cutoff in the Auger spectrum is located at energy below the
predicted GZK cutoff. In any case the shape of the spectrum alone is not 
enough to accept the model (see Section \ref{sec:dip}).

Recently, the very interesting idea of the correlation of $X_{\max}$ and muon 
signals was proposed \cite{correl-th} and the Auger experiment 
\cite{correl-exp} has developed it. Its results exclude a pure proton 
$p$ and $p+He$ mass composition. We argue in Section \ref{sec:correlation} 
that the accuracy of the found correlation is questioned by the anomalous muon 
excess observed in Auger \cite{muon-excess} and TA \cite{TAmuon-excess} 
experiments. With near-future  Auger-Prime measurements the accuracy of 
the measured muon signals $ N_{\mu}$ and $X_{\max}^{\mu}$ will provide 
reliable values of the correlation functions $r(X_{\max},N_{\mu})$ and  
$r(X_{\max},X^{\mu}_{\max})$ together with the conclusions concerning 
mass composition. With these data the accuracy of the correlation analysis 
will undoubtedly reach the needed level providing important informations  
on mass composition.

In order to account for the correlation models of \cite{correl-exp},
considering them as {\em preliminary}, we excluded the pure two-component 
p+He models adding to them CNO nuclei. The CNO component seems also to 
appear in the analysis of \cite{Auger-shape2014} with the EPOS-LHC hadronic 
interaction model. As far as energy spectra are concerned, the model with ratios 
He/p=0.2 and CNO/p=0.1 gives better agreement with  the observed spectrum than the 
two-component p+He model. 

We conclude emphasising that the understanding reached so far on the mass
composition of UHECR is still not conclusive. The observations of mass
composition are still contradictory and cannot exclude a pure light
composition, while the observations of spectra agree fairly well with
such hypothesis. For these reasons the high energy muon program  of
Auger, especially the measurement of $X_{\max}^{\mu} (E)$, will be a
crucial test of the models discussed in this paper.

\vskip 0.3cm
\noindent
We finish with the following note.\\
\indent
This paper is mainly focused on the impact of the Auger analysis
\cite{Auger-shape2014} on the mass composition problem in UHECR.  We
avoided the discussion of some accompanying problems. In particular we
calculated the spectra at  $E \geq 1$~EeV to avoid the discussion
about the transition from galactic to extragalactic cosmic rays. We also
did not include the (possible)  cosmological evolution of the sources,
which can "artificially" improve the agreement of theoretical spectra
with observations. We did not discuss the absence of GZK cutoff in 
all three biggest UHECR detectors: HiRes, Telescope Array and Auger. 
We hope to address these problems in a forthcoming publication.

\vskip 0.3cm
\noindent
The present paper is an updated version of the arXiv paper 1703.08671v1. 

\section*{Acknowledgements} 
We thank Karl-Heinz Kampert, Antonio Villar, Piera Ghia, Michael Unger,
Markus Risse and Alexey Yushkov for valuable remarks and critical comments.
We are grateful to the anonymous Referee for many valuable remarks.

\end{document}